\documentclass[a4paper]{article}
\usepackage{spconf,amsmath,graphicx,cite}

\usepackage{amsbsy}
\usepackage{amsfonts}
\usepackage{amssymb,hhline}
\usepackage[ruled,boxed,boxruled]{algorithm2e}

  \graphicspath{{./}}

 \DeclareGraphicsExtensions{.eps}

\newcommand{\bom}[1]{\boldsymbol{#1}}
\newcommand{\bo}[1]{\mathbf{#1}}
\newcommand{\cov}{\mathrm{Cov}}

\newcommand{\s}{\bo x}  
\renewcommand{\a}{\bo a}  
\newcommand{\es}{x}      
\newcommand{\y}{\bo y} 
 
\renewcommand{\b}{\bo b} 
\renewcommand{\r}{\bo r} 
\newcommand{\ssgn}{\mathrm{\bo{ sign}}} 
\newcommand{\sign}{\mathrm{sign}}

\newcommand{\mm}{\bom \Phi}   
\renewcommand{\S}{\bo X}  
\newcommand{\Y}{\bo Y}  
\newcommand{\E}{\bo R} 
\newcommand{\G}{\bo G} 
\newcommand{\A}{\bo A} 
\newcommand{\B}{\bo B} 
\newcommand{\SNR}{\mathrm{SNR}} 
\newcommand{\I}{\bo I}

\newcommand{\paino}{\sl} 

\newcommand{\Gam}{\Gamma} 
\newcommand{\sig}{\sigma}

\newcommand{\eps}{\bo e} 
\newcommand{\eeps}{e}

\newcommand{\bth}{\bom \theta}

\newcommand{\im}{\jmath} 

\newcommand{\Eps}{\bo E} 

\newcommand{\hop}{\mathrm{H}}

\newcommand{\C}{\mathbb{C}}    

\newcommand{\supp}{\mathrm{supp}}
\newcommand{\rsupp}{\mathrm{rsupp}}

 \renewcommand{\vec}{\mathrm{vec}}

\newcommand{\var}{\sigma^2}    

\newcommand{\beq}{\begin{equation}}
\newcommand{\eeq}{\end{equation}}
\newcommand{\bmat}{\begin{pmatrix}}
\newcommand{\emat}{\end{pmatrix}}
\newcommand{\beqa}{\begin{eqnarray}}
\newcommand{\eeqa}{\end{eqnarray}}

\newcommand{\mb}{\bom \phi}    
\newcommand{\ndim}{M}             
\newcommand{\pdim}{N}             
\newcommand{\kdim}{K}             
\newcommand{\qdim}{Q} 

\newcommand{\MAD}{\sig}   
\newcommand{\SD} {\sigma}  

\title{ Nonparametric Simultaneous Sparse Recovery: an Application to Source Localization}
\name{Esa Ollila}
\address{Aalto University, Dept. of Signal Processing and Acoustics, P.O.Box 13000, FI-00076 Aalto, Finland} 
\begin{document}

\maketitle

\begin{abstract}
We consider multichannel sparse recovery problem  where the objective is to find good recovery of jointly sparse
unknown signal vectors from the given multiple measurement vectors which are different linear combinations of the 
same known elementary vectors.  Many popular greedy or convex algorithms perform poorly under non-Gaussian heavy-tailed noise conditions or in the face of outliers. 
In this paper, we propose the usage of mixed $\ell_{p,q}$ norms on data fidelity (residual matrix)  term 
 and the  conventional $\ell_{0,2}$-norm constraint on the signal matrix to promote row-sparsity. 
We devise a greedy pursuit algorithm based on 
simultaneous normalized  iterative hard thresholding  (SNIHT) algorithm.
Simulation studies highlight the effectiveness of the proposed approaches  to cope with different noise 
environments (i.i.d., row i.i.d, etc) and outliers. Usefulness of the methods are illustrated  in source localization application with sensor arrays. 
\end{abstract}


\section{Introduction} \label{Ollila:sec1}
  
In the  {\paino multiple measurement
vector (MMV) model},  a single measurement matrix is utilized
to obtain multiple measurement vectors, i.e., 
$\y_i = \mm \s_i + \eps_i,$  $i = 1,\ldots,\qdim$  
where 
$
\mm$ 
 is the $\ndim \times \pdim$ known {\it measurement matrix}  and $\eps_i$ 
are the (unobserved) random noise vectors. Typically there are more column vectors   $\mb_{i}$ 
than row vectors $\mb_{(j)}$, i.e., $\ndim<\pdim$ (underdetermined linear model).  It is still possible 
 to recover   the unknown signal vectors $\s_i$, $i=1,\ldots,\qdim$ by assuming  that signals are sparse, i.e., some of the elements are zero.  
In matrix form, the MMV model reads 
$\Y  = \mm \S + \Eps$,
where $\Y = ( \y_1 \, \cdots \,  \y_{\qdim} ) \in \C^{\ndim \times \qdim}$, $\S=( \s_1 \ \cdots \ \s_{\qdim}) \in \C^{\pdim \times \qdim}$ and 
$\Eps =( \eps_1 \ \cdots \ \eps_{\qdim} ) \in \C^{\ndim \times \qdim} $ collect the measurement, the signal  and the error vectors, respectively. 
When $\qdim=1$, the model reduces to standard {\paino compressed sensing (CS) model} \cite{duarte_eldar:2011}. 
Then, rather than recovering the sparse/compressible target signals $\s_i$ separately using standard CS reconstruction algorithms,  
one attempts to simultaneously (jointly) recover all signals.  The key assumption is that locations of nonzero values primarily coincide, 
i.e., signal matrix $\S$ is $\kdim$ rowsparse.   Joint estimation can lead  both to computational advantages and increased reconstruction accuracy  \cite{tropp_etal:2006,tropp:2006,chen_huo:2006,eldar_rauhut:2010,duarte_eldar:2011,blanchard_etal:2014}. 
The objective of multichannel sparse recovery 
problem is finding 
a row sparse approximation of the signal matrix  $\S$ based on knowledge of $\Y$, the measurement matrix $\mm$ and
the sparsity level $\kdim$. 
Applications  
include 
EEG/MEG \cite{duarte_eldar:2011} 
and direction-of-arrival (DOA) estimation  of  sources in array processing \cite{malioutov_etal:2005}. 
  
Most  greedy CS reconstruction algorithms 
have been extended for solving MMV problems. 
These methods, such as simultaneous normalized iterative hard thresholding (SNIHT) algorithm  \cite{blanchard_etal:2014} are guaranteed to perform very well provided that suitable conditions (e.g., incoherence  of $\mm$ and  non impulsive noise conditions) are met. The derived (worst case) recovery bounds depend linearly on $\| \Eps \|_2$, so the methods are not guaranteed to provide accurate reconstruction/approximation  under heavy-tailed non-Gaussian noise.  
In this paper, we consider different $\ell_{p,q}$ mixed norms on data fidelity (residual matrix) 
and   devise a  greedy SNIHT algorithm for obtaining a sparse solution.  
We focus on mixed $\ell_1$ norms as they can provide robust solutions. As will be shown in the sequel,  these methods are then based on {\paino spatial signs} \cite{mot_oja:1995} of the residuals and 
therefore are nonparametric in nature. For an alternative robust approach, see \cite{ollila:2015b}. 

The paper is organized as follows.  In Section~\ref{sec:mnmin}  we formulate a mixed-norm constrained objective function for the MMV problem and motivate the usage   
of $\ell_1$-norm or the mixed $\ell_{2,1}$- and $\ell_{1,2}$-norms.  In  Section~\ref{sec:SNIHT}  we 
formulate the greedy SNIHT algorithm whereas Section~\ref{sec:simul} provides simulation examples  illustrating the improved accuracy of the proposed
methods in various noise conditions and 
signal to noise ratio (SNR) settings. 
Finally, effectiveness of the methods are illustrated  in source localization application with sensor arrays in Section~\ref{sec:array}.


{\bf Notations.} Let $ [n]$ denote the set $\{1, \ldots,n\}$ for $n \in \mathbb{N}^+$. 
For a matrix $\A \in \C^{\ndim \times \pdim}$  and an index set $\Gam$ of cardinality $|\Gam|=\kdim$, we denote 
 by    $\A_{\Gam}$  (resp. $\A_{(\Gam)}$)  the $\ndim \times \kdim$ (resp. $\kdim \times \pdim$) matrix  
restricted to the columns (resp. rows) of $\A$ indexed by the set $\Gam$. 
The $i$th column vector of $\A$ is denoted by $\a_i$ and the hermitian transpose of the  $i$th row vector of $\A$ by $\a_{(i)}$, 
$\A=(\a_1 \ \cdots \ \a_\pdim) = (\a_{(1)} \ \cdots \ \a_{(\ndim)} )^\hop$. 
The {\paino row-support}  of  $\S \in \C^{\pdim \times \qdim}$ is the index set of rows  containing non-zero  elements:
$\rsupp(\S) 
= \{   i \in [\pdim] \, : \: \es_{ij} \neq 0 \,  \mbox{for some $j$} \}.
$
For $p, q \in [1,\infty)$, the {\paino mixed $\ell_{p,q}$ norm} \cite{kowalski2009sparse} of $\S \in \C^{\pdim \times \qdim}$ is defined as
\[
\| \S \|_{p,q} = \bigg( \sum_{i} \bigg(\sum_{j} |\es_{ij}|^p\bigg)^{q/p}  \bigg)^{1/q}  = \left( \sum_{i} \| \s_{(i)} \|_p^q  \right)^{1/q} .
\] 
The mixed norms generalize the usual matrix $p$-norms: if $p=q$, then $\|\S\|_{p,p}=\|\S \|_p$. The $\ell_2$-norm $\|\cdot \|_2$ is called the Frobenius norm and will be denoted shortly as $\| \cdot \|$. 
In the same spirit, the usual Euclidean
norm on vectors is denoted shortly as $\| \cdot \|$. 
The row-$\ell_0$ quasi-norm of a signal matrix $\S$ is the number of nonzero rows, i.e., 
$
\| \S \|_0 = | \ \rsupp(\S) |$. 
The matrix $\S$ is then said to be {\paino $\kdim$-rowsparse} 
if  $\| \S \|_0 \leq \kdim$. 
We use  
$H_\kdim(\cdot)$ to denote the {\paino hard thresholding operator}: 
for a matrix $\S \in \C^{\pdim \times \qdim}$, 
 $H_\kdim(\S)$   retains the elements of the $\kdim$ rows of $\S$ that possess largest  $\ell_2$-norms and set elements of the other rows to zero.    
Notation $\S|_{ \Gam}$ refers to a sparsified version of $\S$ such that the entries
in the rows indexed by set $\Gam$ remain unchanged while all other
rows  have all entries set to $0$. 



\section{Robust mixed norm minimization} \label{sec:mnmin}

Our objective is to recover $\kdim$-rowsparse $\S$ in the MMV model.  For this purpose, we consider the following constrained optimization problem: 
\beq \label{eq:mnmin}
\min_{\S}  c_{p,q} \|  \Y  - \mm \S \|_{p,q}^q  \quad \mbox{subject to} \quad \| \S \|_0  \leq \kdim  \tag{$P_{p,q}$},
\eeq 
where $c_{p,q}$ is an  irrelevant constant used for making notations compact. 
For $p=q$, the problem reduces to conventional  $\ell_p$-norm minimization of the {\paino residual matrix} $\E = \Y  - \mm \S \in \C^{\ndim \times \qdim}$ under row\-sparsity constraint on $\S$.   The well-known problem with $\ell_2$-norm minimization is that it gives a very small weight on small residuals and
a strong weight on large residuals, implying that even a single large outlier can have a large influence
on the obtained solution.  
For robustness, one should utilize $\ell_1$ in mixed norms since it gives larger weights on small residuals and less weight on large residuals. 
In this paper we  consider \eqref{eq:mnmin}  in the cases that  $p,q \in \{1,2\}$. 
The problem \eqref{eq:mnmin} is combinatorial (NP-hard). 
Hence suboptimal reduced complexity reconstruction algorithms
have been proposed. These can be roughly
divided into two classes: convex-relaxation algorithms (e.g., \cite{malioutov_etal:2005,tropp:2006,kowalski2009sparse}) and
greedy pursuit (e.g., \cite{tropp_etal:2006,blanchard_etal:2014}) algorithms. In this paper, we  
devise a greedy simultaneous NIHT (SNIHT) algorithm for the problems  ($P_{1,1}$) and ($P_{2,1}$).  
The case  ($P_{1,2}$) is excluded due to the lack of space,  but our  approach and discussion straightforwardly 
extends for this mixed $\ell_1$ norm  as well. 

In $(P_{1,1})$  problem, one aims to minimize $\|  \Y  - \mm \S \|_1= \sum_{i} \sum_j | y_{ij} - \mb_{(i)}^\hop \s_j | $ 
under sparsity constraint,  so the solution  can be viewed as 
a sparse multivariate least absolute deviation (LAD) regression estimator. The LAD regression (in the real-valued overdetermined linear regression) is well-known to offer robust solution with bounded influence function. 
In the complex case, this approach can be considered optimal when the error terms $\eeps_{ij}$ are i.i.d. with (circular)  complex generalized Gaussian (GG) distribution \cite[Example~Ê4]{ollila_etal:2011} with exponent $s=1/2$.  
It is important to realize that minimization of $\ell_p$-norms in $(P_{p,p}$) implicitly assumes i.i.d.'ness of the error terms.  Since the measurement matrix $\Y$ is in many applications a {{\it space} $\times$ {\it time} matrix as in medical imaging or sensor array applications, 
 the  i.i.d. assumption of the error terms  in time/space is often not valid.   
The benefit of mixed $\ell_1$-norms, such as $\ell_{2,1}$  and $\ell_{1,2}$ considered here is that they introduce couplings \cite{kowalski2009sparse} between the coefficients and offer robustness in case of dependent heavy-tailed errors or outliers. 
When the errors terms have dependencies in time 
and/or  space, then $\ell_{2,1}$  and $\ell_{1,2}$ minimization can offer advantages  over $\ell_1$ or $\ell_2$ norm approaches. As will be shown later, the usage of $\ell_1$-norm or the mixed 
$\ell_1$-norms lead to {\paino non-parametric} approaches that are based on the concept of {\paino spatial sign function} \cite{mot_oja:1995} which in the scalar case ($\es \in \C$) is defined as 
\beq \label{eq:SS} 
\sign(\es)= \begin{cases}   \es/| \es |, &\mbox{for  $\es \neq  0$} \\     0 ,&\mbox{for $\es = 0$} \end{cases}  .
\eeq 
In the vector case, $\ssgn(\s)= \| \s \|^{-1} \s, = 0 $ for $\s \neq \bo 0, = \bo 0$.

\section{Mixed norm SNIHT algorithm} \label{sec:SNIHT}

Iterative hard thresholding is a  {\paino projected gradient descent} 
method that is known to offer efficient and scalable solution for $\kdim$-sparse approximation problem \cite{blumensath_davies:2010}.  
The normalized IHT (NIHT) method updates the estimate of $\S$ by taking steps towards the direction of the negative gradient followed by projection 
onto the constrained space. In our multichannel sparse recovery problem, at $(n+1)$th iteration the SNIHT update is 
\[
\S^{n+1} = H_\kdim \big(\S^{n} \,  + \, \mu^{n+1} \mm^\hop \psi_{p,q}( \Y-\mm \S^{n}) \big)
\]
where $\psi_{p,q}(\E)= \nabla_{\E^*} \| \E \|_{p,q}^q$ is the complex matrix derivative \cite{are_gesbert:2007} with respect to (w.r.t.) $\E^*$,  $\mu^{n+1}>0 $ is the stepsize for the current iteration and $p,q \in \{1, 2\}$.  
For $\ell_2$- and $\ell_1$-norms  the derivatives are easily shown to be 
\[
\psi_{2,2}(\E)=\E \quad \mbox{and} \quad \psi_{1,1} (\E)= \sign(\E)
\] 
respectively, where 
notation $\sign(\E)$ refers to element-wise application of the spatial sign function \eqref{eq:SS}, i.e., $[\sign(\E)]_{ij}=\sign(r_{ij})$.  
For  $(2,1)$ mixed norm, we obtain 
\[
\psi_{2,1}(\E)=  \bmat  \ssgn( \bo r_{(1)}) & \cdots & \ssgn( \bo r_{(\ndim)} ) \emat^\hop ,
\]
that is, the vector spatial sign function is applied row-wise to the residual matrix $\bo R = (\bo r_{(1)} \ \cdots \ \bo r_{(\ndim)})^\hop $. 
Table~\ref{algor:SNIHT} provides the pseudo-code of the greedy SNIHT algorithm for the problem $(P_{p,q})$, which we call SNIHT$(p,q)$ algorithm for short. 
Note that SNIHT$(2,2)$ corresponds to the conventional SNIHT studied in \cite{blanchard_etal:2014} and 
in \cite{blumensath_davies:2010} for $\qdim=1$ case. 

\begin{algorithm}
\caption{SNIHT$(p,q)$ algorithm}  \label{algor:SNIHT}
\DontPrintSemicolon
\SetKwInOut{Input}{input}\SetKwInOut{Output}{output}
\SetKwFunction{Support}{InitSupport}
\SetKwInOut{Init}{initialize}
\SetAlgoNlRelativeSize{-1}
\SetNlSkip{0.4em}
\Input{$\Y$, $\mm$, sparsity $\kdim$, mixed norm indices $(p,q)$ }
\Output{$(\S^{n+1},\Gam^{n+1})$  estimates of $\S$ and $\rsupp(\S)$} 

\Init{$\S^0=\bo 0$, $\mu^0=0$, $\Gam^0 = \emptyset$, $n=0$.}
 
 \BlankLine 
\nl $\Gam^0   = \rsupp\big(H_K( \mm^\hop\psi_{p,q}( \Y) ) \, \big)$     

\While{ halting criterion false}{\label{InRes1} 

\nl $\E^{n}_\psi = \psi_{p,q}(\Y-\mm \S^{n})$  

\nl $\G^n  = \mm^\hop \E^n_\psi$ 

\nl $\mu^{n+1}  =$ {\tt CompStepsize}($\Phi,\G^n, \Gam^n,\mu^n,p,q$)   

\nl $\S^{n+1} = H_\kdim (\S^{n} \,  + \, \mu^{n+1} \G^n )$  

\nl $\Gam^{n+1} = \rsupp(\S^{n+1})$  

\nl $n=n+1$

 }
\end{algorithm}

We now describe the  {\tt CompStepsize}  function which computes the stepsize update $\mu^{n+1}$ in Step~4. 
Following the approach in \cite{blumensath_davies:2010}, assuming that we have identified the correct support at $n$th iteration, then we may look for a  stepsize update $\mu^{n+1}$ as the minimizer of $  \| \Y - \mm \S\|_{p,q}^q$   for the gradient ascent direction $\S^n + \mu \G^n|_{ \Gamma^n}$. 
Thus we  find $\mu>0$ as the minimizer of the convex function 
\begin{align} \label{eq:Lmu}
 \left\| \Y - \mm \big( \S^n + \mu \G^n|_{ \Gamma^n} \big) \right\|_{p,q}^q = \left\|  \E^n - \mu \bo B^n \right\|_{p,q}^q  
\end{align}
where $\E^n=\Y - \mm \S^n$ and  $\B^n = \mm_{\Gam^n} \G_{(\Gamma^n)}$ When $p=q$ this reduces to minimizing  a simple linear  regression estimation problem, $\min_\mu \| \bo r- \mu \bo b\|_p^p$,  where  the response is $\bo r=\vec(\E^n)$ and the  predictor is $\bo b=\vec(\B^n)$. 
Thus when using $p=q=2$ as in conventional SNIHT \cite{blanchard_etal:2014}, 
the minimizer of \eqref{eq:Lmu} is easily found to be $\mu^{n+1} = \| \G_{(\Gam^n)}^n  \|^2 / \| \mm_{\Gam^n}   \G_{(\Gam^n)}^n \|^2$. However, for the robust estimators that 
we are interested in, i.e., when using $(p,q)=(1,1)$ and $(p,q)=(2,1)$, a minimizer of  \eqref{eq:Lmu} can not be found in closed-form. In the $(p,q)=(1,1)$ case, it is easy to show that 
the solution $\mu$ verfies the following fixed point (FP) equation  $\mu  = H(\mu)$, where 
\[
H(\mu)=\Big( \sum_{i,j}   | \tilde r_{ij} |^{-1} | b_{ij} |^2 \,\Big)^{-1}  \sum_{i,j}  | \tilde r_{ij} |^{-1} \mathrm{Re}(b_{ij}^* r_{ij}),   
\] 
and $\tilde \E =  \E^n - \mu  \bo B^n = (\tilde r_{ij})$  depends also on the unknown $\mu$. Then, instead of choosing the next update $\mu^{n+1}$ as the minimizer of \eqref{eq:Lmu} which could be found 
by running the FP iterations $\mu_i= H(\mu_{i+1})$ for $i=0,1,\ldots$  until convergence  (with  initial value $\mu_0>0$), 
we use a 1-step FP iterate which corresponds to a single iteration with initial value of iteration given by the previous stepsize $\mu^{n}$. In other words, in Step~Ê4, we set $\mu^{n+1}=H(\mu^n)$.  
In our simulation studies we noticed that this  1-step FP iterate often  gave a very good approximation of the true solution (within 3 decimal accuracy). In case we use $(p,q)=(2,1)$, it is easy to show that 
the solution $\mu$ verifies the FP equation  $\mu  = H^\star(\mu)$, where
\[
H^\star(\mu)=\Big( \sum_{i}  \| \tilde \r_{(i)} \|^{-1} \| \b_{(i)} \|^2 \, \Big)^{-1}  \sum_{i}  \| \tilde \r_{(i)} \|^{-1} \mathrm{Re}(\b_{(i)}^\hop \r_{(i)})  
\] 
and the same approach, i.e.,  $\mu^{n+1}=H^\star(\mu^n)$, is used for computing the stepsize update.

\section{Simulation studies} \label{sec:simul}

Next we illustrate the  usefulness of the  methods  in  a variety of noise environments 
and SNR levels.  Also, the effect of number of measurement vectors $\qdim$ on recovery probability will be illustrated. 
The elements of $\mm$ are drawn from $\C \mathcal{N}(0,1)$ distribution and the columns are unit-norm normalized. 
The coefficients of  $\kdim$ non-zero row vectors of $\S$ have  equal amplitudes $\sigma_x = |\es_{ij}|=1$ $\forall i \in \Gamma$, $j=1,\ldots,\qdim$ and uniform phases, i.e., $\mathrm{Arg}(\es_{ij}) \sim 
Unif(0,2\pi)$. The support 
$\Gamma=\mathrm{supp}(\S)$ is randomly chosen from $\{1,\ldots,\pdim\}$  
without replacement for each trial. We define the (generalized) {\it signal to noise ratio (SNR)} as
$\mathrm{SNR}(\sig) = 10 \log_{10}(\sigma_x^2/\sig^2) = -20 \log_{10} \sigma$ 
which depends on the scale parameter $\sigma$ of the error distribution. 
For robustness purposes, we will study the performance in i.i.d. complex circular $t$-distributed noise with $\nu$ degrees of freedom (d.o.f.), $e_{ij} \sim \C t_\nu (0,\sig^2)$, when 
$\nu \leq 5$ and the scale parameter is  $\var= \mathrm{Med}_{F_{\eeps}}(|\eeps_{ij}|^2)$.   This is an example of  a heavy-tailed distribution with $\nu=1$ corresponding  to Cauchy distribution. Also note that at the limit $\nu \to \infty$ one obtains the complex Gaussian distribution.

As performance measures of sparse signal recovery, we use both the (observed) {\paino mean squared error}
$
\mathrm{MSE}(\hat \S) = \frac{1}{L\qdim} \sum_{\ell =1}^L \big\| \hat \S^{[\ell]} - \S^{[\ell]} \big\|^2
$
and the  empirical {\paino probability of exact recovery},   
$
\mbox{PER}  \triangleq \frac{1}{L}\sum_{\ell=1}^L  \mathrm{I}\!\big( \hat \Gamma^{[\ell]} = \Gamma^{[\ell]}\big), 
$
where $\mathrm{I}(\cdot)$ denotes the indicator function, 
$\hat{\S}^{[\ell]}$ and $\hat \Gamma^{[\ell]} =\rsupp(\hat \S^{[\ell]})$ denote the  estimate of the $\kdim$-sparse signal $\S^{[\ell]}$  and the signal support $\Gamma^{[\ell]}$  for the $\ell$th Monte Carlo (MC) trial,  respectively. 
In all simulation settings described below, all the reporeted figures are averages over $L=2000$ MC trials, the length of the signal is $\pdim=512$, the number of measurements is $\ndim=256$, 
and the sparsity level is $\kdim=8$.  The number of measurement vectors is $\qdim = 16$ unless otherwise stated. 

\begin{figure}[!t]
\centerline{ \includegraphics[width=0.27\textwidth]{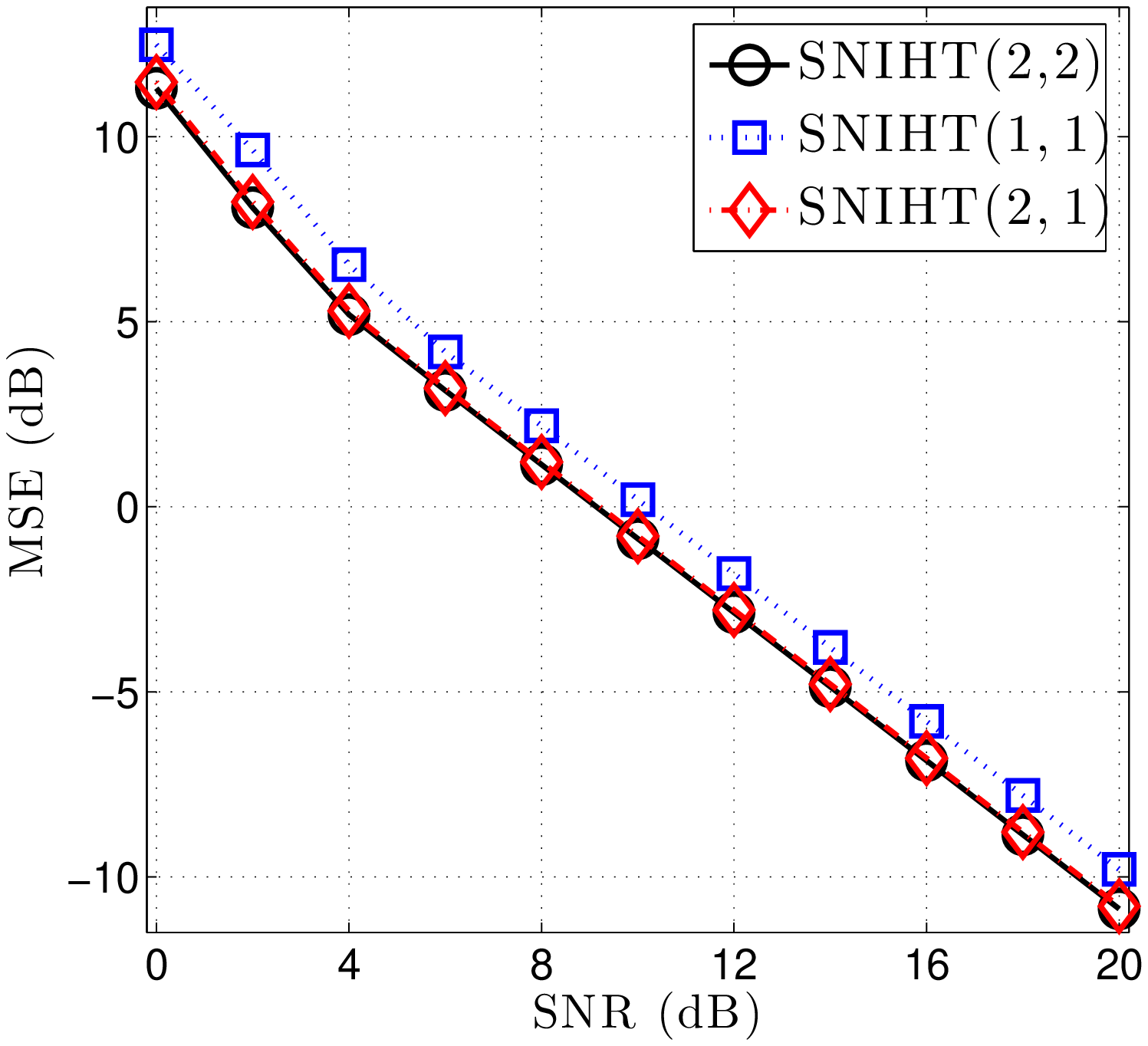}\hspace{-0.5cm}  \includegraphics[width=0.27\textwidth]{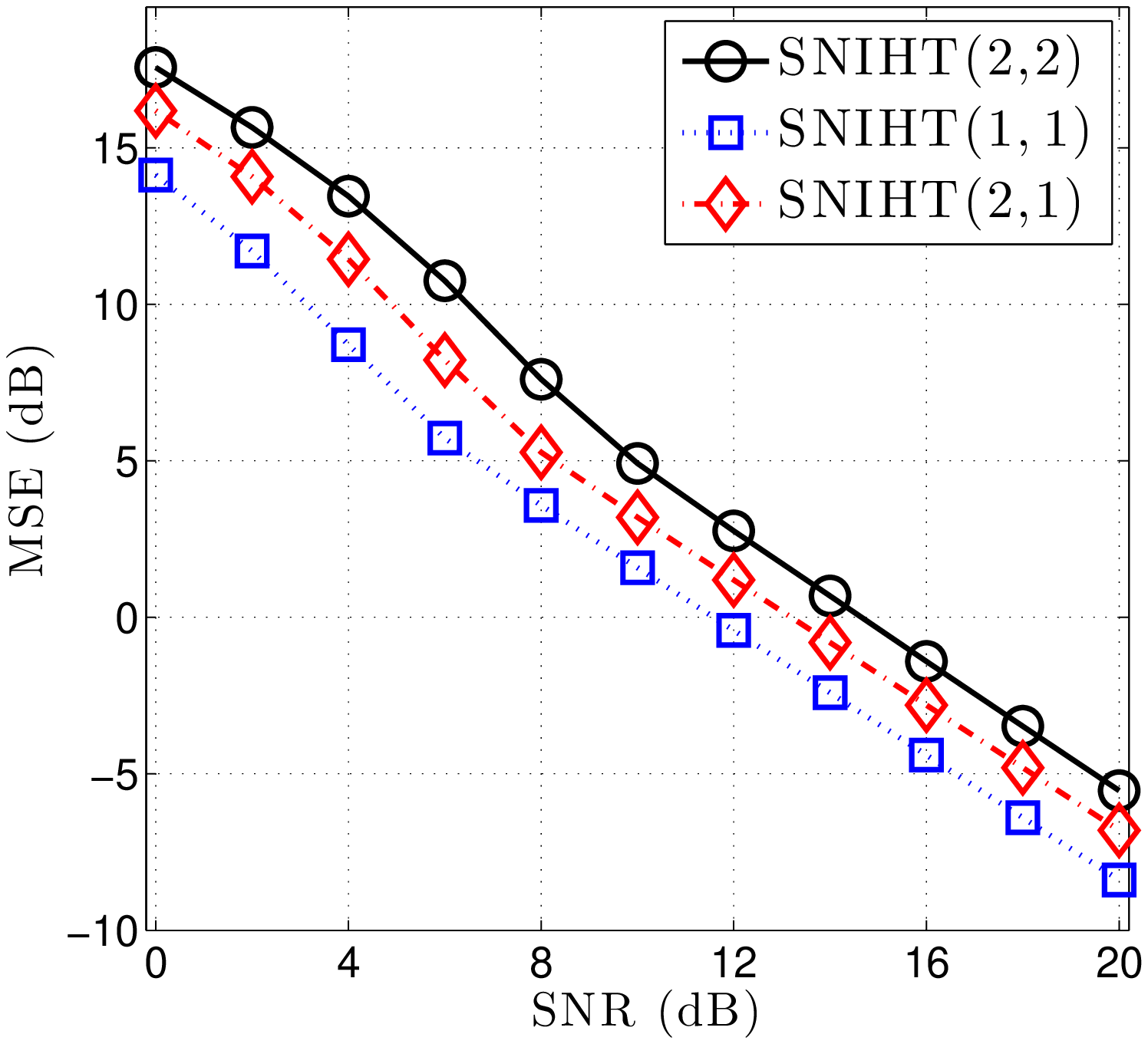}}
\centerline{(a)  \hspace{3.6cm} (b) }
\vspace{-0.2cm}
\caption{Average MSE of SNIHT$(p,q)$Ê{} methods as a function of $\mathrm{SNR}$ 
in (a) $\C \mathcal N(0,\SD^2)$   
noise and (b)  $\C t_3(0,\SD^2)$ noise.} \label{Ollila:fig2}
\end{figure}

\begin{table}[!t]
\begin{center}
{\small
  \begin{tabular}{ |   l  |   c  |  c | c | c  | c | c | c  |  c | c |   }
  \hline
  &  \multicolumn{8}{c|}{SNR (dB) } \\ \hhline{~--------}
  &       2  &   4   &  6   &  8   & 10   & 12  &  14  &  16 \\  \hline
SNIHT$(2,2)$  &        0  &     0    &       .6  &      .61    &    .94    &    .99   &     .99 &      1.0 \\
SNIHT$(1,1)$ &        0  &    .25  &       .91 &      1.0  &     1.0 &    1.0  &     1.0 &      1.0 \\
 SNIHT$(2,1)$  &       0 &    .02   &    .38   &       .96  & 1.0   & 1.0      &  1.0    &    1.0   \\ \hline
 \end{tabular}
}
\end{center} 
\caption{PER rates in $\C t_3(0,\SD)$ distributed noise as a function of SNR (dB). System parameters were $(\ndim,\pdim,\kdim,\qdim)=(256,512,8,16)$.} \label{tab:v3}
\end{table}

Figure~\ref{Ollila:fig2}(a) depicts the MSE as a function of SNR in i.i.d. circular Gaussian noise, $e_{ij} \sim \C \mathcal N(0,\sig^2)$, 
where $\var =  \mathbb{E}[|\eeps_{ij}|^2]$. 
As expected, the conventional SNIHT$(2,2)$ has the best performance, but SNIHT$(2,1)$ suffers a negligible $0.07$ dB loss, whereas  SNIHT$(1,1)$  attain  $1.07$ dB performance loss. Note that $\SNR = 6$ dB 
is the cutline for which all methods had full PER rate ($= 1$). From 4 dB the PER rate declines and reaches 0 at $\SNR = 0$ dB for all of the methods.   

Next we study the performance in $t$-distributed noise with  $\nu=3$ d.o.f. Note that $\C t_3(0,\SD)$ distribution has a finite variance so we can expect that also SNIHT$(2,2)$ can still work reliably in this setting. Figure~\ref{Ollila:fig2}(b) which depict the MSE vs SNR illustrates severe degradation in reconstruction performance 
for the SNIHT$(2,2)$. This is further illustrated in Table~\ref{tab:v3} which provides the PER rates for the considered SNIHT$(p,q)$ methods. Note that 
the decline of PER rate starts much earlier  for the conventional SNIHT than for the robust methods.

 Figure~\ref{fig:tsim1}(a) depicts the MSE of the methods  in  $t$-distributed noise of  $\mathrm{SNR}(\MAD)=10$ dB and  d.o.f. $\nu$ varying in $\nu \in [1,5]$.  
We observe that SNIHT$(1,1)$ has the best performance as it  retains low MSE for all values of $\nu$.  This is in deep contrast to  SNIHT$(2,2)$ which starts an exponential increase at $\nu \leq 3$, 
reaching sky-high MSE levels in Cauchy noise ($\nu=1$).  The PER rates 
in Table~\ref{tab:tdist} further illustrates the remarkable performance of the robust methods. 
Note that SNIHT$(1,1)$  is able to maintain full PER rates for all values of $\nu$, whereas  SNIHT$(2,2)$ fails completely for $\nu<3$.   

The usefulness of joint recovery becomes more pronounced at low SNR's, where multiple measurements can dramatically improve on the recovery by exploiting the joint information.
This is illustrated in our next simulation set up,  where d.o.f. $\nu$ of the $t$-distributed noise is fixed at $\nu=3$  and the SNR is $10$ dB. 
Figure~\ref{fig:tsim1}(b) depicts the PER rates for increasing number of measurement vectors $\qdim$. 
As can be seen, the  PER rate increases as a function of $\qdim$ from poor $14$\% (when $\qdim=2$) to near full 100\% recovery (when $\qdim = 6$) when using SNIHT$(1,1)$ method.  Again, 
SNIHT$(2,1)$ is slightly behind  in performance to SNIHT$(1,1)$. Conventional SNIHT$(2,2)$ is drastically behind the robust methods, reaching highest 96.6\% rate when $\qdim=18$. This is again in deep contrast with near $100$\% PER obtained by SNIHT$(1,1)$  method only with  $\qdim=6$ samples.

\begin{figure}[!t]
\centerline{ \includegraphics[width=0.27\textwidth]{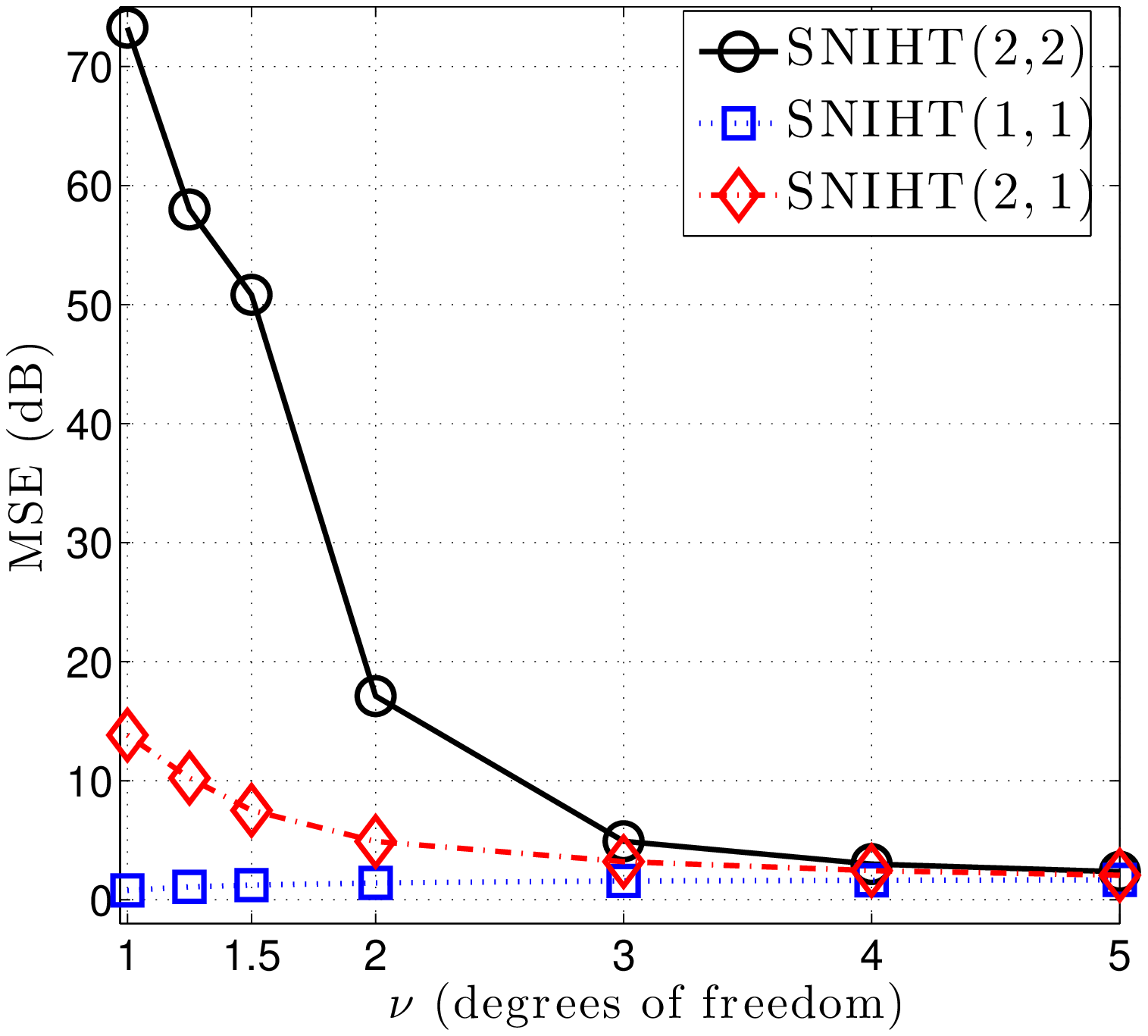}  \hspace{-0.5cm} \includegraphics[width=0.27\textwidth]{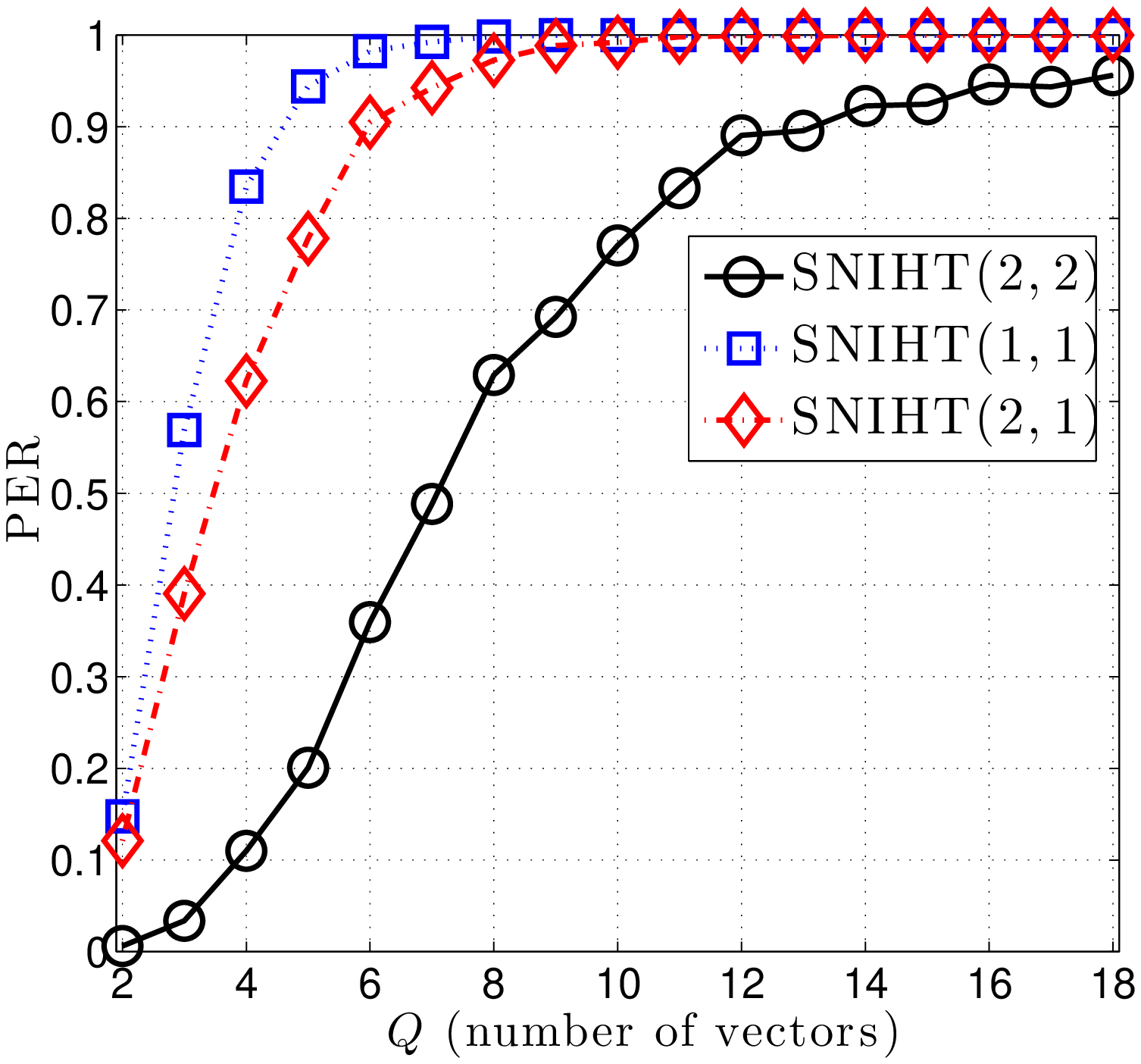} }
\centerline{(a)  \hspace{3.3cm} (b) }
\caption{(a) MSE of SNIHT$(p,q)$  methods in $\C t_\nu(0,\MAD^2)$ noise  as a function of $\nu$; (b) Empirical PER rates of SNIHT$(p,q)$Ê{} methods as a function of $\qdim$ in $\mathcal \C t_3(0,\SD^2)$
noise. In both setting, the SNR was $\SNR(\MAD)= 10$ dB. 
 } \label{fig:tsim1}
\end{figure}

\begin{table}
  \begin{center} 
    \begin{tabular}{ |     c | c | c | c | c | c | c | c | c | }
\hline 
    SNIHT        &  \multicolumn{8}{c|}{Degrees of freedom $\nu$ } \\ 
    \hhline{~--------}
 $(p,q)$ &  1     &  1.25  & 1.5   & 1.75 &  2         &  3    &    4     & 5 \\ \hline
  $(2,2)$                 & 0      &    0       &  0    &     0    & .04   &  .94 &    .99 &    1.0 \\ 
 $(2,1)$     &     0   &  .07 &    .55  &    .90  &  .98 &    1.0  &  1.0   & 1.0 \\
 $(1,1)$   	       & 1.0   &   1.0   &  1.0   &  1.0  &   1.0   &  1.0  & 1.0  &   1.0  \\  
               \hline
               \end{tabular}
                              \end{center}

               \vspace*{-0.3cm}
               \caption{PER rates in i.i.d. $\C t_\nu(0,\sig^2)$ noise for different d.o.f. $\nu$ and 
 $\mathrm{SNR}(\MAD)=10 $ dB.  System parameters were $(\ndim,\pdim,\kdim,\qdim)=(256,512,8,16)$.
   } \label{tab:tdist}
    \label{table:rawpatient}
\end{table}

\section{Applications to Source Localization} \label{sec:array} 

Consider a sensor array consisting of $\ndim$ sensors that receives $\kdim$ narrowband incoherent farfield plane-wave sources from a point source ($\ndim>\kdim$). 
At discrete time $t$,  the {\paino array output} (snapshot) $\y(t) \in \C^m$ is a weighted linear 
combination of the signal waveforms $\s(t) = (\es_1(t), \ldots,\es_\kdim(t))^\top$ 
corrupted by additive noise $\eps(t) \in \C^\ndim$,  
$\y(t) = \A(\bom \theta) \s(t) + \eps(t)$,  
where  $\A=\bo A(\bom \theta)$ is the $\ndim \times \kdim$ {\paino steering matrix} para\-met\-rized by the vector $\bom \theta=(\theta_1,\ldots,\theta_\kdim)^\top$ 
of  (distinct) unknown DOA's  of the sources. 
Each column vector $\a(\theta_i)$, called  the {\paino steering vector}, 
represents a point in known array manifold $\a(\theta)$. 
We assume  that the number of sources $\kdim$ is known.  
 
As in \cite{malioutov_etal:2005},   we cast the source localization problem as a multichannel sparse recovery problem as follows. 
We construct an overcomplete $\ndim \times \pdim$ steering matrix $\A(\tilde \bth)$, where $\tilde \bth=(\tilde \theta_{1}, \ldots, \tilde \theta_{\pdim})^\top$ represents a sampling
grid of all source locations of interest. Suppose that $\tilde \bth$ contains the true DOA's $\theta_i$, $i=1,\ldots,\kdim$.  
In this case the measurement matrix $\Y=\bmat \y(t_1) & \cdots & \y(t_\qdim) \emat \in \C^{\ndim \times \qdim}$ 
 consisting of snapshots at time instants $t_1,\ldots,t_\qdim$ can be {\it exactly} modelled as MMV model 
 in which the signal matrix $\S \in \C^{\pdim \times \qdim}$ is $\kdim$-rowsparse matrix, whose $\kdim$ non-zero row vectors
correspond to  source signal sequences. 
Thus finding the DOA's of the sources is equivalent to identifying the support $\Gam=\supp(\S)$. 
Since the steering matrix $\A(\tilde \bth)$ is known, we can use SNIHT methods  to 
identify the support.

We assume that $\kdim=2$ independent (spatially and temporally) complex circular Gaussian source signals of equal power $\sigma^2_x$ arrive on uniform linear array (ULA) of $\ndim=20$ sensors  with half a wavelength inter-element spacing from DOA's $\theta_1=0^o$ and $\theta_2=8^o$. 
In this case, the array manifold is $ \a(\theta) = ( 1,e^{-\im \pi  \sin(\theta)},\cdots,e^{-\im \pi (\ndim-1) \sin(\theta) })^\top$. 
The noise matrix $\Eps \in \C^{\ndim \times \qdim}$ has i.i.d. row vectors, each row vector $\eps_{(i)}$ having complex $\qdim$-variate inverse Gaussian compound Gaussian (IG-CG) distribution 
\cite{ollila_etal:2012b} with shape parameter 
$\lambda=0.1$ and covariance matrix $\cov(\eps_{(i)})= \I_\qdim$.   Note that the covariance of the snapshot is $\cov(\y(t_i))= \sigma_x^2 \A(\bth)  \A(\bth)^\hop + \I_\ndim$, so we may 
use the popular MUSIC method to localize the sources. In other words, we search for the $\kdim=2$ peaks of the MUSIC pseudospectrum in the grid. 
We use a uniform grid $\tilde \bth$ on $[-90, 90]$ with 2$^o$ degree spacing, thus containing the true DOA's. In Step~1 of SNIHT$(p,q)$ algorithm, we locate the $K$ largest peaks 
of rownorms of $\mm^\hop\psi_{p,q}( \Y)$ instead of taking $\Gam^0$ as indices of $\kdim$ largest rownorms of $\mm^\hop\psi_{p,q}( \Y)$. 

We then identify the support (which gives the DOA estimates) for all the methods over 1000 MC trials  
and compute the PER rates and the relative frequency of DOA estimates in the grid. Full  PER rate $=1$  implies that the support $\Gam$  correctly identified the true DOA's 
in all MC trials. Such a case is shown in upper plot of Figure~\ref{Ollila:fig3b} for the SNIHT$(1,1)$  and SNIHT$(2,1)$ when the number of snapshots is $\qdim=50$ and the SNR is $-10$ dB. 
The PER rates of SNIHT$(2,2)$ and MUSIC  were considerably lower, $0.81$ and $0.73 $, respectively. 
Next we keep other parameters fixed, but decrease the SNR to  -20 dB. In this case, the MUSIC method fails completely and provides nearly a uniform frequency on the grid. This is illustrated  in lower plot of Figure~\ref{Ollila:fig3b}. Note that the proposed  
robust methods, SNIHT$(1,1)$ and SNIHT$(2,1)$, provide high peaks on the correct DOA's. The PER rates of 
SNIHT$(2,1)$,  SNIHT$(1,1)$, SNIHT$(2,2)$ and MUSIC were    $0.70$,   $0.64$, $0.11$ and   $0.05$, respectively. Hence the mixed $\ell_1$-norm method SNIHT$(2,1)$  has the best recovery performance.  
In conclustion, robust sparse recovery methods can offer considerable improvements in  performance 
when the  measurement environment  is challenging (low SNR,  small $\qdim$)
  
\begin{figure}[!t]
\centerline{ \includegraphics[width=0.54\textwidth]{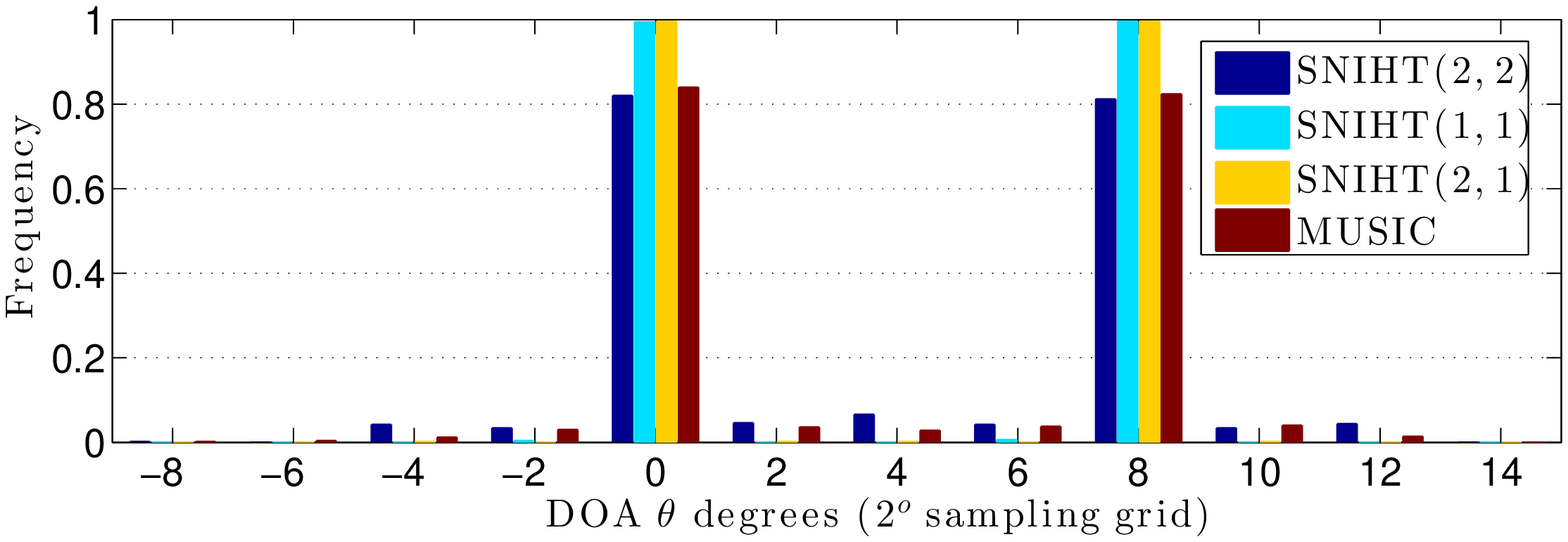}}
\centerline{ \includegraphics[width=0.54\textwidth]{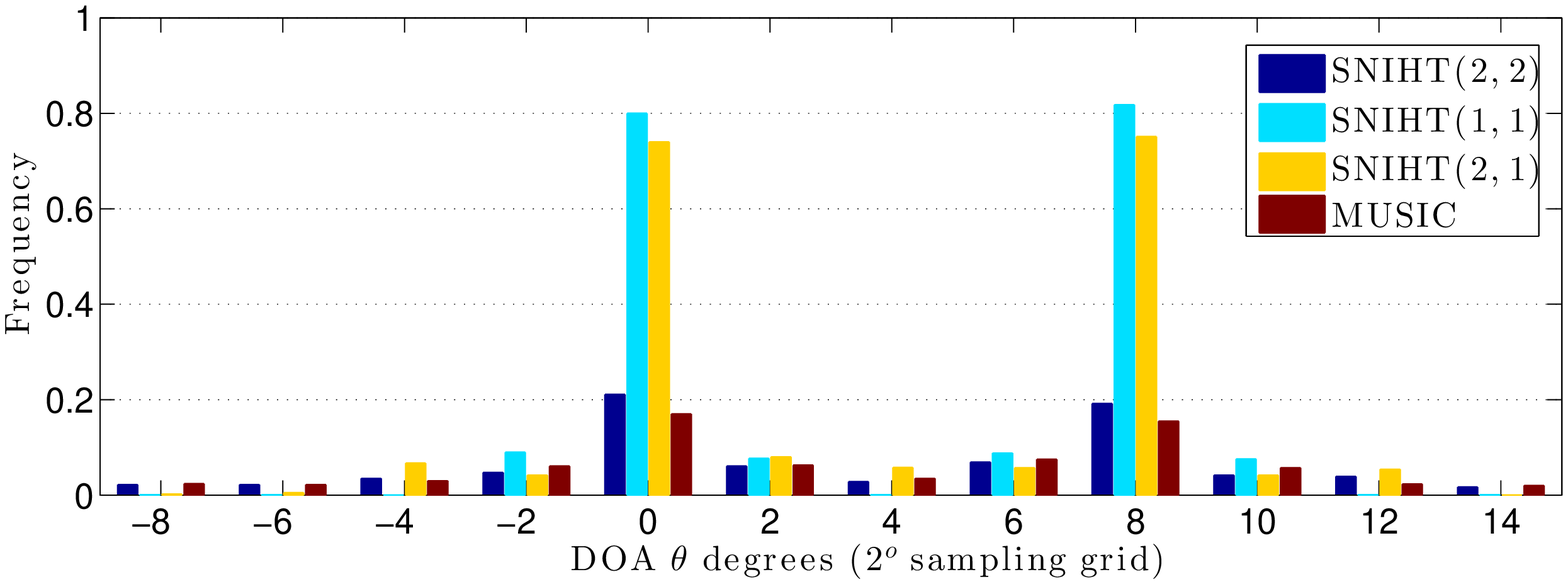}}
 \vspace*{-0.3cm}
\caption{Relative frequency of DOA estimates. 
Two equal power  Gaussian sources arrive from  
 DOA  $0^o$ and $8^o$ and 
the noise has i.i.d. row vectors following IG-CG distribution with covariance matrix $\I$ and shape $\lambda=0.1$. 
 $\SNR(\MAD)=- 10$ dB (upper plot) and $\SNR(\MAD)=- 20$ dB (lower plot). 
}  \label{Ollila:fig3b}
\end{figure}


{\small 
\bibliographystyle{IEEEbib}
\bibliography{../../IEEE/bibtex/IEEEabrv,../../mybib/STATabrv,../../mybib/STATbib,../../mybib/ENGbib}
}
\end{document}